\documentclass{article}[11pt]
\usepackage{a4wide}
\usepackage{bm}
\usepackage{bbm}
\usepackage{epsfig,graphics}
\usepackage{amsmath,amssymb,amsfonts}
\usepackage{color}
\usepackage{epstopdf}
\usepackage{slashed}
\usepackage[
colorlinks=true,
linkcolor=black,
breaklinks=true,
urlcolor=blue,
citecolor=green]{hyperref}

\newcommand{\be}{\begin{equation}}
\newcommand{\ee}{\end{equation}}
\newcommand{\ba}{\begin{eqnarray}}
\newcommand{\ea}{\end{eqnarray}}

\newcommand{\al}{&\!\!\!}

\newcommand{\order}[1]{\mathcal{O}\left(#1\right)}

\begin{document}

\title{Disentangling the hadronic molecule nature of the $P_c(4380)$
pentaquark-like structure}
\date{\today}

\author{Chao-Wei Shen$^{1,2,}$\footnote{Email address:
      \texttt{shencw@itp.ac.cn} }~ ,
      Feng-Kun Guo$^{1,}$\footnote{Email address:
      \texttt{fkguo@itp.ac.cn} }~, Ju-Jun Xie$^{3,1,}$\footnote{Email address:
      \texttt{xiejujun@impcas.ac.cn} }~ and
      Bing-Song Zou$^{1,2,}$\footnote{Email address:
      \texttt{zoubs@itp.ac.cn} }
        \\[2mm]
      {\it\small$^1$Key Laboratory of Theoretical Physics, Institute
      of Theoretical Physics,}\\
      {\it\small  Chinese Academy of Sciences, Beijing 100190,China}\\
      {\it\small$^2$University of Chinese Academy of Sciences (UCAS), Beijing 100049, China} \\
      {\it\small$^3$Institute of Modern Physics, Chinese Academy of Sciences, Lanzhou 730000, China}
}

\maketitle

\begin{abstract}

We demonstrate that the relative ratio of the decays of hidden charm
pentaquark-like structure $P^+_c(4380)$ to $\bar{D}^* \Lambda_c^+$
and $J/\psi p$ are very different for $P^+_c$ being $\bar{D}
\Sigma_c^*(2520)$ or $\bar{D}^* \Sigma_c(2455)$ molecule states.
While the partial width of the $\bar{D} \Sigma_c^*(2520)$ molecule
to the  $\bar{D}^* \Lambda_c^+$ is much larger, by one order of
magnitude, than that to the $J/\psi p$, the $\bar{D}^*
\Sigma_c(2455)$ molecule shows a different pattern. Our analysis
shows that the $\bar{D} \Sigma_c^*$ bound state ansatz is more
reasonable than the $\bar{D}^* \Sigma_c$ one to explain the broad
$P_c(4380)$ structure. We suggest to search for the $P_c(4380)$ in
the $\bar{D}^* \Lambda_c^+$ system, which can be used to disentangle
the nature of the $P^+_c(4380)$ structure.

\end{abstract}

\medskip

\newpage

\section{Introduction}

Exploration of the exotic baryons that have more than three constituent quarks
is an important issue in hadron physics.
Recently, observation of two hidden-charm pentaquark-like structures
$P^+_c(4380)$ and $P^+_c(4450)$ in the $J/\psi p$ invariant mass distribution in
the process of $\Lambda^0_b \to J/\psi p K^-$ decay was reported by the LHCb
Collaboration~\cite{Aaij:2015tga}. The values of the masses and widths from the fit
with the Breit--Wigner parameterization are $M_{P_c(4380)} = (4380 \pm 8 \pm
29)$~MeV, $\Gamma_{P_c(4380)} = (205 \pm 18 \pm 86)$~MeV, $M_{P_c(4450)} =
(4449.8 \pm 1.7 \pm 2.5)$~MeV, and $\Gamma_{P_c(4450)} = (39 \pm 5 \pm 19)$ MeV,
with spin-parity $J^P$ being either $3/2^\pm$ or $5/2^\mp$. Possible existence
of such pentaquark states with hidden charm has already been
predicted~\cite{Wu:2010jy,Yuan:2012wz,Yang:2011wz,Xiao:2013yca,Uchino:2015uha}
prior to the experimental observation. In the earliest
prediction~\cite{Wu:2010jy},  a $\bar{D}^* \Sigma_c(2455)$ S-wave bound state
with $J^P=3/2^-$ was predicted to be around 4412 MeV with $J/\psi N$ as its
largest decay mode, in the framework of the meson-baryon coupled channel unitary
approach with the local hidden gauge formalism. In this approach, the
$t$-channel vector meson exchange dominance is assumed for the $\bar{D}^* \Sigma_c$
interaction. Taking into account of other meson exchanges, the mass of the
predicted $\bar{D}^* \Sigma_c$ S-wave bound state could be shifted by $\pm 40$
MeV~\cite{Yang:2011wz}. Considering coupled channel effects with $\bar{D}
\Sigma_c^*$ and  $\bar{D}^* \Sigma_c^*$ channels, three $J^P=3/2^-$ pentaquark
states were predicted to be around 4334 MeV, 4417 MeV and 4481 MeV, mainly
coupled to $\bar{D} \Sigma_c^*$, $\bar{D}^* \Sigma_c$ and $\bar{D}^*
\Sigma_c^*$, respectively~\cite{Xiao:2013yca}. Therefore both $P^+_c(4380)$ and
$P^+_c(4450)$ could be the predicted $\bar{D} \Sigma_c^*$ and $\bar{D}^*
\Sigma_c$ states. The predicted masses for genuine pentaquark states with both
negative and positive parity~\cite{Yuan:2012wz} suffer large model dependence,
but also cover the observed masses of the two $P^+_c$ structures.

After the observation of the two $P^+_c$ structures, many
theoretical works have been triggered, see for example,
Refs.~\cite{Chen:2015loa,Chen:2015moa,Roca:2015dva,Mironov:2015ica,He:2015cea,
Huang:2015uda,Burns:2015dwa,Guo:2015umn,Meissner:2015mza,
Liu:2015fea,Lebed:2015tna,Wang:2015epa,Scoccola:2015nia,Zhu:2015bba,Shimizu:2016rrd},
proposing various explanations for these structures. Among them, it
was suggested that the observed structures could be due to
kinematical triangle singularities~\cite{Guo:2015umn,Liu:2015fea},
the possibility of which needs to be examined by future experiments.
If these two $P^+_c$ structures correspond to two particle states,
since they sit close to the mass thresholds of the $\bar{D}
\Sigma_c^*$ and $\bar{D}^* \Sigma_c$ at 4387~MeV and 4461~MeV,
respectively, a popular explanation for them is still either
$S$-wave $\bar{D} \Sigma_c^*(2520)$ or $\bar{D}^* \Sigma_c(2455)$
molecular
states~\cite{Chen:2015loa,Chen:2015moa,Roca:2015dva,Mironov:2015ica,He:2015cea,Huang:2015uda,Burns:2015dwa}
with $J^P = \frac{3}{2}^-$, roughly consistent with previous
predictions~\cite{Wu:2010jy,Yang:2011wz,Xiao:2013yca} but with
parameters tuned to reproduce the observed masses of $P_c^+$
structures. However, the observed decay width of the $P^+_c(4380)$
state is about a few times larger than the predicted
one~\cite{Wu:2010jy,Xiao:2013yca}. Also the LHCb experiment claims
that the two states have opposite parity, which is against that both
states are S-wave molecules of $\bar D^{(*)}\Sigma_c^{(*)}$ to have
spin-parity of $3/2^-$.

In this work, we want to make an estimate of the partial decay widths of the
$P_c(4380)$ into the $\bar{D}^* \Lambda_c^+$ and $J/\psi p$ assuming it be a
hadronic molecular state. We will point out that the previous
prediction~\cite{Xiao:2013yca} for the decay width of $\bar{D} \Sigma_c^*(2520)$
bound state underestimates the contribution of the $\bar{D}^* \Lambda_c^+$ decay
by more than an order of magnitude due to its assumption of vector meson
exchange dominance. Rather than the $J/\psi p$ mode, the dominant decay mode for
a $\bar{D} \Sigma_c^*(2520)$ bound state should be $\bar{D}^* \Lambda_c^+$ due
to $t$-channel pion exchange.  We demonstrate that the relative ratio of the
decays of hidden charm $P^+_c$ pentaquark states to $\bar{D}^* \Lambda_c^+$ and
$J/\psi p$ are very different for $P^+_c$ to be $\bar{D} \Sigma_c^*(2520)$ or
$\bar{D}^* \Sigma_c(2455)$ molecular states. While the $\bar{D}
\Sigma_c^*(2520)$ molecule decays dominantly into the $\bar D^*\Lambda_c^+$, the
$\bar{D}^* \Sigma_c(2455)$ molecule has a larger branching fraction for the
decay into the $J/\psi p$.
Therefore, were the $P_c$ structures hadronic molecular states, future
measurement of this ratio can help us to pin down their nature.
The unexpected large decay width of the $P^+_c(4380)$ can get a natural
explanation if it is a $\bar{D} \Sigma_c^*(2520)$ molecule.

This article is arranged as follows. In the next section, we present
the theoretical framework of our calculation. In
Sect.~\ref{sec:results}, the numerical results and some discussions
are presented.

\section{Theoretical framework} \label{sec:formalism}

\begin{figure}[htbp]
\centering
\includegraphics[scale=0.6]{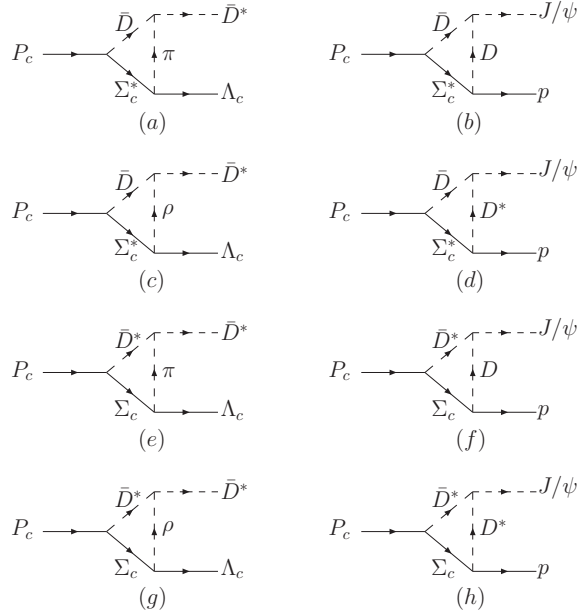}
\caption{Diagrams representing the decays of the $P^+_c(4380)$ state
to $\bar{D}^* \Lambda^+_c$ and $J/\psi p$ as $\bar{D}
\Sigma_c^*(2520)$ molecule (a-d) or $\bar{D}^* \Sigma_c$ molecule
(e-h).
\label{Fig:feyndiag}}
\end{figure}

Among the two observed structures, the existence of the narrow
$P_c(4450)$, no matter what it is, is affirmative from the data for
the $J/\psi p$ invariant mass distribution. However, introducing the
second structure, the $P_c(4380)$, does not seem that necessary
since there is still a discrepancy between the best fit with two
$P_c$ structures and the data in the right shoulder of the peak in
the $J/\psi p$ invariant mass distribution. Yet, a recent
phenomenological analysis of the data affirms the necessity of
introducing the $P_c(4380)$~\cite{Roca:2016tdh}. The nominal mass of
the $P_c(4380)$ is just 7 MeV below the $\bar{D} \Sigma_c^*$
threshold and 81 MeV below $\bar D^*\Sigma_c$ threshold. It seems
more natural to be a $\bar{D} \Sigma_c^*$ dominant
molecule~\cite{Xiao:2013yca,He:2015cea,Shimizu:2016rrd}. However the
possibility to be a deeply bounded $\bar{D}^* \Sigma_c$ state cannot
be excluded~\cite{Chen:2015loa}. Here, we assume the $P^+_c(4380)$
exists with the properties reported by the LHCb Collaboration, and
study its decays to the two final states $\bar{D}^* \Lambda^+_c$ and
$J/\psi p$ with the assumption that it is a bound state of $\bar{D}
\Sigma_c^*(2520)$ (type I) or $\bar{D}^* \Sigma_c(2455)$ (type II).
These two decays can proceed through triangular diagrams as shown in
Fig~\ref{Fig:feyndiag}. Since we only aim at making a rough
estimate, which is sufficient for the conclusion, we only consider
the exchange of lightest possible mesons. This means that we will
consider the one-pion-exchange, as well as the one-rho-exchange in
previous work~\cite{Wu:2010jy}, between the charmed baryons and
anti-charmed mesons, and the exchange of ground state pseudoscalar
and vector charmed mesons, which are related to each other via heavy
quark spin symmetry, for the decays into the $J/\psi p$.


In order to evaluate the decay amplitudes of the diagrams shown in
Fig.~\ref{Fig:feyndiag}, we need the structure of the involved interaction
vertices which can be described by means of the following effective
Lagrangian~\cite{Molina:2008jw,MuellerGroeling:1990cw},
\begin{eqnarray}
{\cal L}_{PPV} &=& g_{PPV} \phi_P(x) \partial_{\mu} \phi_P(x) \phi_V^{\mu}(x),\\
{\cal L}_{VVP} &=& g_{VVP} i \varepsilon_{\mu \nu \alpha \beta} \partial^{\mu} \phi_V^{\nu}(x)
\partial^{\alpha} \phi_V^{\beta}(x) \phi_P(x),\\
{\cal L}_{VVV} &=& g_{VVV} i \left[\partial_{\mu} \phi_{V \nu}(x) -
\partial_{\nu} \phi_{V \mu}(x)\right] \phi_V^{\mu}(x) \phi_V^{\nu}(x),\\
{\cal L}_{BPB^*} &=& g_{BPB^*} \left[\bar{\psi}_{B^* \mu}(x)\psi_B(x) +
\bar{\psi}_B(x)\psi_{B^* \mu}(x)\right]  \partial^{\mu} \phi_P(x),\\
{\cal L}_{BVB^*} &=& g_{BVB^*} i \big[\bar{\psi}_{B^* \nu}(x) \gamma^5
\gamma_\mu \psi_B(x) - \bar{\psi}_B(x) \gamma^5 \gamma_\mu  \psi_{B^* \nu}(x)\big]
\left[\partial^{\mu} \phi_V^\nu(x)- \partial^{\nu} \phi_V^\mu(x)\right],\\
{\cal L}_{BBP} &=& g_{BBP} \bar{\psi}_B(x) i \gamma^5 \psi_B(x) \phi_P(x),\\
{\cal L}_{BBV} &=& g_{BBV} \big[ \bar{\psi}_B(x) \gamma_\mu \psi_B(x)
\phi_V^\mu(x)+2f_{BBV}  \bar{\psi}_B(x) \sigma_{\mu \nu} \psi_B(x)\left(\partial^{\mu}
\phi_V^\nu(x)-\partial^{\nu} \phi_V^\mu(x)\right) \big],
\end{eqnarray}
where $P, V, B, B^*$ denote pseudoscalar, vector meson, octet and
decuplet baryon, respectively.
The coupling constants $g_{DD^*\pi}$,  $g_{\Sigma_c\Lambda_c\pi}$ and
$g_{\Sigma_c^*\Lambda_c\pi}$ can be determined from the experimental data of the
decay widths of the $D^*$, $\Sigma_c$ and $\Sigma_c^*$, respectively. The
extracted values of $g_{\Sigma_c\Lambda_c\pi}$ and $g_{\Sigma_c^*\Lambda_c\pi}$
fulfills very well the relation predicted by heavy quark spin symmetry (HQSS).
The coupling constant $g_{D^*D^*\pi}$ can be related to the value of
$g_{DD^*\pi}$ by heavy quark spin symmetry.
The other coupling constants cannot be measured directly. Since we
only aim at making a rough estimate of the partial decay widths, we
take model
values~\cite{Chen:2013bha,Oh:2000qr,Oh:2007jd,Can:2012tx,Ronchen:2012eg,Aliev:2011kn,
Aliev:2010ev,Aliev:2010yx,Aliev:2010su,Garzon:2015zva,Guo:2010ak,Gao:2010ve,Kim:2015ita}
for them which are listed in Table~\ref{table:para}.

\begin{table}[htbp]
    \begin{center}
\caption{\label{table:para} The values of coupling constants
involved in the
calculation~\cite{Chen:2013bha,Oh:2000qr,Oh:2007jd,Can:2012tx,Ronchen:2012eg,Aliev:2011kn,
Aliev:2010ev,Aliev:2010yx,Aliev:2010su,Garzon:2015zva,Guo:2010ak,Gao:2010ve,Kim:2015ita}.}
    \begin{tabular}{c c c c}
        \hline
        \hline
        Coupling constant & Value & Coupling constant & Value \\
        \hline
        $g_{D D^* \pi}$ & 6.3 & $g_{D D J/\psi}$ & 7.4 \\
        \hline
        $g_{D D^* \rho}$ & $2.8 ~{\rm GeV}^{-1}$ & $g_{D D^* J/\psi}$ & $2.5
        ~{\rm GeV}^{-1}$  \\
        \hline
        $g_{D^* D^* \pi}$ & $6.3 ~{\rm GeV}^{-1}$ & & \\
        \hline
        $g_{D^* D^* \rho}$ & 5.9 & $g_{D^* D^* J/\psi}$ & 8.0 \\
        \hline
        $g_{\Sigma^*_c \Lambda_c \pi}$ & $7.4~{\rm GeV}^{-1}$ & $g_{\Sigma^*_c
        ND}$ & $6.5~{\rm GeV}^{-1}$ \\
        \hline
        $g_{\Sigma^*_c \Lambda_c \rho}$ & $10.0~{\rm GeV}^{-1}$ & $g_{\Sigma^*_c
        ND^*}$ & $2.9~{\rm GeV}^{-1}$ \\
        \hline
        $g_{\Sigma_c \Lambda_c \pi}$ & $ 9.3 $ & $g_{\Sigma_c ND}$ & $ 2.7 $
        \\
        \hline
        $g_{\Sigma_c \Lambda_c \rho}$ & $ 0.4 $ & $g_{\Sigma_c N D^*}$ & $ 3.0$
        \\
        \hline
        $f_{\Sigma_c \Lambda_c \rho}$ & $8.1~{\rm GeV}^{-1}$ & $f_{\Sigma_c N D^*}$ & $0.6~{\rm GeV}^{-1}$ \\
        \hline
        \hline
    \end{tabular}
    \end{center}

\end{table}

While for those interaction vertices including the spin-$3/2$ $P_c(4380)$ state,
we use the Lorentz covariant orbital-spin ($L$-$S$) scheme as illustrated in
Ref.~\cite{Zou:2002yy}. With this scheme, we can easily write down the effective
Lagrangians as
\begin{eqnarray}
{\cal L}_{P_c(\frac{3}{2}^-) \Sigma_c \bar{D}^*} &=& g_{P_c \Sigma_c
\bar{D}^*}
\bar{\Sigma}_c P_{c \mu} \bar{D}^{* \mu} + {\rm H.c.},  \\
{\cal L}_{P_c(\frac{3}{2}^-) \Sigma_c^* \bar{D}} &=& g_{P_c
\Sigma_c^* \bar{D}} \bar{\Sigma}_c^{* \mu} P_{c\mu} \bar{D} + {\rm
H.c.},
\end{eqnarray}
where $P_c$ is the pentaquark fields with $J^P = 3/2^-$.
Here we have assumed that the $P_c$ is an $S$-wave hadronic molecular state of
either $\bar D^*\Sigma_c$ or $\bar D\Sigma_c^*$.
When we are only interested in the ratio between the partial widths to $\bar
D^*\Lambda_c$ and $J/\psi p$ of a given hadronic molecule structure, either
$\bar D^*\Sigma_c$ or $\bar D\Sigma_c^*$, the coupling constant gets cancelled.

Combining the Lagrangians and propagators given above together, we
can easily get the decay amplitudes for the process shown in
Fig.~\ref{Fig:feyndiag}, and the expressions are given in Appendix A.

The loop integrals in the amplitudes are ultraviolet (UV) divergent, which means
that we need counterterms to absorb the divergence. Here, in order to be able to make an
estimate we will neglect the counterterms and simply use a Gaussian
regulator with the cutoff taking values in a large range. For the explicit form
of the regulator, we take the one used in
Refs.~\cite{Faessler:2008zza,Chen:2015igx,Lu:2016nnt}:
\begin{equation}
\label{corr_fun}
\Phi_{P_c}(q_E^2/\Lambda^2) \equiv \exp( - q_E^2/\Lambda^2)\,,
\end{equation}
where $q_{E}$ is the Euclidean Jacobi momentum.


The partial decay width of the two-body decay of the $P_c(4380)$
state in its rest frame is given by
\begin{eqnarray}
{\rm d}\Gamma = \frac{1}{32 \pi^2} \overline{|{\cal M}|^2}
\frac{|\mathbf{p_2}|}{M^2} {\rm d}\Omega,
\end{eqnarray}
where $M$ is the mass of the $P_c(4380)$, while $\textbf{p}_2$
is the $\Lambda_c$ (or $p$) three-momentum in the rest frame
of the $P_c(4380)$. The averaged squared amplitude
$\overline{|{\cal M}|^2}$ can be obtained from
\begin{eqnarray}
\overline{|{\cal M}|^2}   &=&   \frac{1}{4} \sum_{s_{P_c}}
\sum_{s_{\Lambda_c}, s_{\bar{D}^*}} |{\cal M}|^2,
\end{eqnarray}
for the $P_c(4380) \to \bar{D}^* \Lambda_c$ decay, with

\begin{eqnarray}
{\cal M} &=& {\cal M}_a + {\cal M}_c, ~~{\rm for ~type ~I}, \\
{\cal M} &=& {\cal M}_e + {\cal M}_g, ~~{\rm for ~type ~II},
\end{eqnarray}
and
\begin{eqnarray}
\overline{|{\cal M}|^2}   &=&   \frac{1}{4} \sum_{s_{P_c}}
\sum_{s_{p}, s_{J/\psi}} |{\cal M}|^2,
\end{eqnarray}
for $P_c(4380) \to J/\psi p$ decay, with
\begin{eqnarray}
{\cal M} &=& {\cal M}_b + {\cal M}_d, ~~{\rm for ~type ~I}, \\
{\cal M} &=& {\cal M}_f + {\cal M}_h, ~~{\rm for ~type ~II},
\end{eqnarray}
where ${\cal M}_{a,b,c,d,e,f,g,h}$ are given in Appendix~\ref{app:amp}.

\section{Results and discussions} \label{sec:results}

\begin{figure}[htbp]
\centering
\includegraphics[scale=0.6]{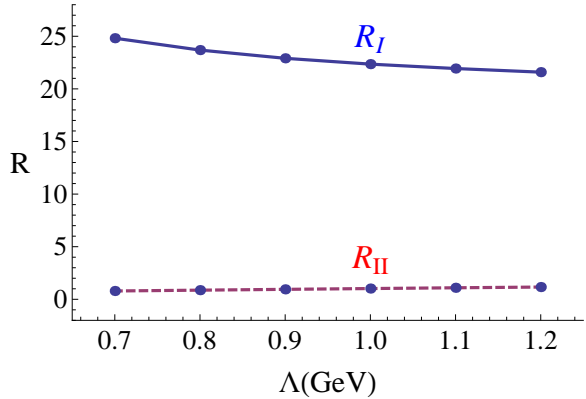}
\caption{The $\Lambda$ dependence of $R$ for the cases of type I and
II. \label{Fig:result}}
\end{figure}

The partial decay width is proportional to $g^2_{P_c \Sigma_c^*
\bar{D}}$ and $g^2_{P_c \Sigma_c \bar{D}^*}$ for the cases of type I
and type II, respectively, which are canceled in the calculation of
the ratio $R$ defined as
\begin{eqnarray}
R_\text{I} = \frac{\Gamma(P_c(4380) \to \bar{D} \Sigma_c^* \to \bar{D}^*
\Lambda_c)} {\Gamma(P_c(4380) \to \bar{D} \Sigma_c^* \to J/\psi p)} ,
\end{eqnarray}
for the case of type I and
\begin{eqnarray}
R_\text{II} = \frac{\Gamma(P_c(4380) \to \bar{D}^* \Sigma_c \to \bar{D}^*
\Lambda_c)} {\Gamma(P_c(4380) \to \bar{D}^* \Sigma_c \to J/\psi p)} ,
\end{eqnarray}
for the case of type II. Using the values of the coupling constants given in
Table~\ref{table:para}, we show diagrammatically the $\Lambda$ dependence of $R$
for the case of type I and type II in Figs.~\ref{Fig:result}.

One sees that the dependence of both ratios on the cutoff is rather weak. The
partial decay width of the $P_c$ into the $\Lambda_c\bar D^*$ is much larger
than that into the $J/\psi p$ for the type I hadronic molecule, while the
situation is different for type II. Because the $P_c$ structures were observed
by the LHCb Collaboration in the $J/\psi p$ invariant mass distribution, our
results show that the $P_c$ should be copiously produced in $\Lambda_c\bar
D^*$ and thus can be easily searched for by reconstructing events for
$\Lambda_c$ and $\bar D^*$ if it is a type~I hadronic molecule.
Therefore, this ratio can be employed to tell the nature of the
$P_c$ resonances in the future experiments, such as experiments at LHCb,
the $\gamma p$ experiments at JLab~\cite{Huang:2013mua}, or the $\pi p$
experiments at JPARC~\cite{Lu:2015fva}.

It is a firm conclusion that the partial width of $P_c(4380)\to \bar
D^*\Lambda_c$ for the $P_c(4380)$ being a $\bar{D} \Sigma_c^*$ hadronic molecule
is much larger than the $\bar{D}^* \Sigma_c $ hadronic molecular case. This
conclusion does not depend on any unknown coupling constant, and is analyzed in
details using the nonrelativistic formalism taking heavy quark spin
symmetry into account in Appendix~\ref{app:nr}.

We also find that the
ratio $R$ is insensitive to the mass of the $P_c$
in the range between 4.36~GeV and 4.50~GeV which covers the locations of both
LHCb $P_c$ structures. Yet, we need to notice that because of the mass,
the $P_c(4450)$, located 10~MeV below the $\bar D^*\Sigma_c$ threshold, cannot
be a $\bar D \Sigma_c^*$ bound state.


\begin{figure}[htbp]
\centering
\includegraphics[scale=1]{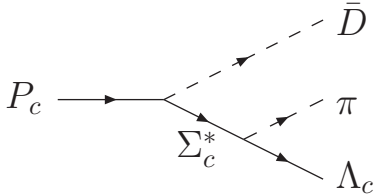}
\caption{Diagram representing the decay  $P^+_c(4380)\to\bar{D} \pi \Lambda^+_c$
for the $P_c(4380)$ being a $\bar D\Sigma_c^*$ hadronic molecule.
\label{Fig:3bodydiag}}
\end{figure}
A hadronic molecule with unstable constituents can decay naturally through
the decays of its constituents. However, the widths of $\Sigma_c^{(*)}$ are
small, which leads to small three-body decay widths for the $P_c$. For instance,
the three-body decay $P_c \to \bar{D} \pi \Lambda_c$ shown in
Fig.~\ref{Fig:3bodydiag} leads to a width of only 7.3~MeV,
much
smaller than the reported width of the $P_c(4380)$. Here we evaluated
the value of the coupling $g_{P_c \Sigma_c^* \bar{D}}$ using~\cite{Weinberg:1965zz,Baru:2003qq}
\begin{eqnarray}
{g^2} = \frac{4\pi}{4 M m_2}  \frac{(m_1+m_2)^{5/2}} {(m_1 m_2)^{1/2}}
\sqrt{32\epsilon} , \label{eq:coco}
\end{eqnarray}
where $M$, $m_1$ and $m_2$ are the masses of $P_c$, $\bar{D} (\bar{D}^*)$ and $\Sigma_c^* (\Sigma_c)$, respectively,
and $\epsilon$ is the binding energy,
which is valid for an $S$-wave shallow bound state.
Here we have introduced the factor $1/(4 M m_2)$ to account for the
normalization of fermion fields in comparison with the formula used in, e.g.
Ref.~\cite{Guo:2008zg}. If we take the mass of the $P_c$ as 4.38~GeV, then
$g_{P_c \Sigma_c^* \bar{D}} = 1.3$.
The large value of $R_I$ makes possible that the $\bar D\Sigma_c^*$ molecule
decays dominantly into the $\bar D^*\Lambda_c$ rather than the three-body
tree-level decay mode. We can make an order-of-magnitude estimate of
$\Gamma(P_c(4380) \to \bar{D}^* \Lambda_c)$ for type~I hadronic molecule. Taking
the cutoff to be in the range between 0.7~GeV and 1.2~GeV, which reflects the
intrinsic model dependence because of the UV divergence of the loop integrals,
the partial width in question could be as large as
$\mathcal{O}(100~\text{MeV})$. The nonrelativistic formalism with a Gaussian
form factor as described in Appendix~\ref{app:nr} leads to the same conclusion.

It is worthy to mention that our results also depend on the values of
those coupling constants shown in Table~\ref{table:para}, and some of them are
obtained from flavor $SU(4)$. Fortunately, as shown in Figs.~\ref{Fig:result},
the magnitude of $R$ for type I and type II are different by
more than an order of magnitude, hence even if these values only present a rough
estimate of the real values of the coupling constants, our main conclusion
should still be valid.
The large decay branching ratio of the $\bar{D} \Sigma_c^*$ molecule to
$\bar{D}^* \Lambda_c$ results in a much larger decay width than that of the
$\bar{D}^* \Sigma_c$ molecule.
Furthermore, the nominal mass of the $P_c(4380)$ is just a few MeV below the
$\bar{D} \Sigma_c^*$ threshold. These properties makes more plausible to explain
the $P_c(4380)$ as a $\bar{D} \Sigma_c^*$ hadronic molecule than a  $\bar{D}^*
\Sigma_c$ one.

In summary, we have studied the decays of hidden charm pentaquark
$P^+_c(4380)$ state to $\bar{D}^* \Lambda_c^+$ and $J/\psi p$, assuming that its
quantum numbers are $J^P=3/2^-$, under the hadronic molecular assumption of
either $\bar D\Sigma_c^*$ or $\bar D^*\Sigma_c$.
The two decays can be described by means of the triangle diagrams
where the two constituents of the $P^+_c$ can exchange the
pseudoscalar and vector mesons leading to the $J/\psi p$ or
$\bar{D}^* \Lambda_c^+$ final states. We estimate the ratio of these
two decay modes.
The results show that its value is
sensitive to the ansatz of whether the $P^+_c$ is a $\bar{D}
\Sigma_c^*(2520)$ bound state or a $\bar{D}^* \Sigma_c(2455)$ bound state.
According to our calculation, if the $P_c(4380)$ is a $\bar{D} \Sigma_c^*$
bound state, it would have a much larger branching ratio to the $\bar{D}^*
\Lambda_c$ than that to the $J/\psi p$.
And the situation is different  if the $P_c(4380)$ is a $\bar{D}^*
\Sigma_c$ bound state. As a result, the $\bar{D} \Sigma_c^*$ bound state ansatz
is more reasonable than the $\bar{D}^* \Sigma_c$ one to explain the broad
$P_c(4380)$ structure. We
suggest to search for the $P_c(4380)$ in the $\bar{D}^* \Lambda_c^+$ system,
which can be used to disentangle the nature of the $P^+_c(4380)$ structure.

\section*{Acknowledgments}

We thank Yu-Bing Dong, Ulf-G. Mei\ss{}ner and Jia-Jun Wu for
useful discussions. This work is supported by the National Natural Science
Foundation of China under Grants: No.~11475227, No.~11035006, No.~11121092, and
No.~11261130311 (CRC110 by DFG and NSFC), by the Chinese Academy of Sciences
under Project No.~KJCX2-EW-N01, and by the Thousand Talents Plan for Young
Professionals. It is also supported by the Open Project Program of State Key Laboratory of
Theoretical Physics, Institute of Theoretical Physics, Chinese
Academy of Sciences, China (No.~Y5KF151CJ1).

\newpage

\begin{appendix}

\section{Decay amplitudes}
\label{app:amp}

The amplitudes involved in the calculation are

\begin{eqnarray}
{\cal M}_a &=& g_{P_c \Sigma_c^* \bar{D}} g_{D^* D \pi} g_{\pi
\Lambda_c \Sigma^*_c} \int_{-\infty}^{\infty} \frac{{\rm d}^4 q}{(2
\pi)^4}\Phi_{P_c}(q_E^2/\Lambda^2) \bar{u}(p_2,s_{\Lambda_c})
(p_2-q)_\mu G^{\mu \nu}_{\Sigma^*_c} (q) u_{\nu}(p_1,s_{P_c}) \times
\nonumber \\
&& G_{\bar{D}} (p_1 -q)
G_{\pi} (q - p_2) (p_1 + p_2 -
2q)^{\lambda} \epsilon^*_{\lambda}(p_1-p_2, s_{\bar{D}^*}), \label{eq:Ma} \\
{\cal M}_b &=& g_{P_c \Sigma_c^* \bar{D}} g_{D D J/\psi} g_{D N
\Sigma^*_c} \int_{-\infty}^{\infty} \frac{{\rm d}^4 q}{(2
\pi)^4}\Phi_{P_c}(q_E^2/\Lambda^2) \bar{u}(p_2,s_p) (p_2-q)_\mu
G^{\mu \nu}_{\Sigma^*_c} (q) u_{\nu}(p_1,s_{P_c}) \times \nonumber \\
&& G_{\bar{D}} (p_1 -q)
G_{\bar{D}} (q - p_2) (p_1 + p_2 - 2q)^{\lambda}
\epsilon^*_{\lambda}(p_1-p_2, s_{J/\psi}),      \label{eq:Mb} \\
{\cal M}_c &=& g_{P_c \Sigma_c^* \bar{D}} g_{D^* D \rho} g_{\rho
\Lambda_c \Sigma^*_c} \int_{-\infty}^{\infty} \frac{{\rm d}^4 q}{(2
\pi)^4}\Phi_{P_c}(q_E^2/\Lambda^2) \bar{u}(p_2,s_{\Lambda_c}) \gamma_5
((\slashed p_2-\slashed q) g_{\mu \nu} - \gamma_\mu
(p_2-q)_\nu)  \times
\nonumber \\
&& G^{\nu \alpha}_{\Sigma^*_c} (q) u_{\alpha}(p_1,s_{P_c})G_{\bar{D}} (p_1 -q)
G_{\rho}^{\mu \tau} (q - p_2) \varepsilon_{\theta \tau \phi \lambda}
(p_2 -q)^\theta (p_1 - p_2)^{\phi}
\epsilon^{* {\lambda}}(p_1-p_2, s_{\bar{D}^*}), \label{eq:Mc} \\
{\cal M}_d &=& g_{P_c \Sigma_c^* \bar{D}} g_{D D^* J/\psi} g_{D^* N
\Sigma^*_c} \int_{-\infty}^{\infty} \frac{{\rm d}^4 q}{(2
\pi)^4}\Phi_{P_c}(q_E^2/\Lambda^2) \bar{u}(p_2,s_p)  \gamma_5
((\slashed p_2-\slashed q) g_{\mu \nu} - \gamma_\mu
(p_2-q)_\nu)  \times
\nonumber \\
&& G^{\nu \alpha}_{\Sigma^*_c} (q) u_{\alpha}(p_1,s_{P_c}) G_{\bar{D}} (p_1 -q)
G_{D^*}^{\mu \tau} (q - p_2) \varepsilon_{\theta \tau \phi \lambda}
(p_2 -q)^\theta (p_1 - p_2)^{\phi}
\epsilon^{* \lambda}(p_1-p_2, s_{J/\psi}),      \label{eq:Md} \\
{\cal M}_{e} &=& g_{P_c \Sigma_c \bar{D}^*} g_{D^* D^* \pi} g_{\pi
\Lambda_c \Sigma_c} \int_{-\infty}^{\infty} \frac{{\rm d}^4
q}{(2\pi)^4} \Phi_{P_c}(q_E^2/\Lambda^2)\bar{u}(p_2,s_{\Lambda_c})
\gamma_5 G_{\Sigma_c}(q) u_\mu(p_1,s_{P_c})\times \nonumber \\
&& G^{\mu \nu}_{\bar{D}^*}(p_1-q)
\varepsilon_{\alpha \beta \kappa \nu} (p_1 - q)^\alpha
(p_1 - p_2)^\kappa G_{\pi}(q - p_2) {\epsilon}^{* \beta}(p_1-p_2, s_{\bar{D}^*}),  \label{eq:Me}  \\
{\cal M}_{f} &=& g_{P_c \Sigma_c \bar{D}^*} g_{D D^* J/\psi} g_{D N
\Sigma_c} \int_{-\infty}^{\infty} \frac{{\rm d}^4
q}{(2\pi)^4}\Phi_{P_c}(q_E^2/\Lambda^2) \bar{u}(p_2,s_{p})
\gamma_5 G_{\Sigma_c}(q) u_\mu(p_1,s_{P_c}) \times \nonumber \\
&& G^{\mu \nu}_{\bar{D}^*}(p_1-q)
\varepsilon_{\alpha \beta \kappa \nu} (p_1 - q)^\alpha
(p_1 - p_2)^\kappa G_{\bar{D}}(q - p_2) {\epsilon}^{* \beta}(p_1-p_2, s_{J/\psi}), \label{eq:Mf} \\
{\cal M}_{g} &=& g_{P_c \Sigma_c \bar{D}^*} g_{D^* D^* \rho} g_{\rho
\Lambda_c \Sigma_c}
 \int_{-\infty}^{\infty} \frac{{\rm d}^4 q}{(2\pi)^4}\Phi_{P_c}(q_E^2/\Lambda^2) \bar{u}(p_2,s_{\Lambda_c}) (\gamma_\beta - f_{\rho \Lambda_c \Sigma_c}
(\gamma_\beta (\slashed p_2-\slashed q)  \nonumber \\
&& - (\slashed p_2-\slashed q) \gamma_\beta)) G_{\Sigma_c}(q) u_\mu(p_1,s_{P_c}) G^{\mu \nu}_{\bar{D}^*}(p_1-q)
G^{\tau \beta}_{\rho}(q - p_2) (g_{\tau \nu}(2q-p_1-p_2)_\alpha  \nonumber \\
&& + g_{\alpha \tau}(2p_2-q-p_1)_\nu + g_{\nu \alpha}(2p_1-p_2-q)_\tau)
{\epsilon}^{* \alpha}(p_1-p_2,s_{\bar{D}^*}), \label{eq:Mg} \\
{\cal M}_{h} &=& g_{P_c \Sigma_c \bar{D}^*} g_{D^* D^* J/\psi}
g_{D^* N \Sigma_c} \int_{-\infty}^{\infty} \frac{{\rm d}^4
q}{(2\pi)^4} \Phi_{P_c}(q_E^2/\Lambda^2)\bar{u}(p_2,s_p)
(\gamma_\beta - f_{D^* N \Sigma_c}
(\gamma_\beta (\slashed p_2-\slashed q) \nonumber \\
&& - (\slashed p_2-\slashed q) \gamma_\beta)) G_{\Sigma_c}(q) u_\mu(p_1,s_{P_c}) G^{\mu \nu}_{\bar{D}^*}(p_1-q)
G^{\tau \beta}_{\bar{D}^*}(q - p_2) (g_{\tau \nu}(2q-p_1-p_2)_\alpha  \nonumber \\
&& + g_{\alpha \tau}(2p_2-q-p_1)_\nu  + g_{\nu \alpha}(2p_1-p_2-q)_\tau)
{\epsilon}^{* \alpha}(p_1-p_2,s_{J/\psi}), \label{eq:Mh}
\end{eqnarray}
where $u_{\nu}$ and $u$ are dimensionless Rarita-Schwinger and Dirac
spinors, respectively, while $\epsilon^*_{\lambda}(p_1-p_2,
s_{\bar{D}^*})$ ($\epsilon^*_{\lambda}(p_1-p_2, s_{J/\psi})$) is the
$\bar{D}^*$ ($J/\psi$) polarization vector. Here $p_1$, $p_2$ and
$q$ are the momentums of $P_c(4380)$, $\Lambda_c$ (or $p$) and
$\Sigma_c^*$ (or $\Sigma_c$), respectively. Besides, the $s_{P_c}$,
$s_{\bar{D}^*}$, $s_{\Lambda_c}$, $s_{J/\psi}$, and $s_p$ are
polarization variables for $P_c(4380)$, $\bar{D}^*$, $\Lambda_c$,
$J/\psi$, and $p$, respectively. $G_{\pi/D} (q)$, $G^{\mu
\nu}_{D^*} (q)$, $G_{\Sigma_c} (q)$, and $G^{\mu
\nu}_{\Sigma^*_c} (q)$ are the propagators for the $\pi$, ($D$ or
$\bar{D}$), $D^*$ or $\bar{D}^*$, $\Sigma_c$, and $\Sigma^*_c$, respectively,
which are
\begin{eqnarray}
G_{\pi/D} (q) &=& \frac{1}{q^2 - m^2_{\pi/D}}, \\
G^{\mu \nu}_{D^*} (q) &=& \frac{-g^{\mu \nu} + {q^\mu
q^\nu}/q^2}{q^2 - m^2_{D^*}},\\
G_{\Sigma_c} (q) &=& \frac{\slashed q + m_{\Sigma_c}}{q^2 -
m^2_{\Sigma_c}}, \\
G^{\mu \nu}_{\Sigma^*_c} (q) & =&  \frac{\slashed q +
m_{\Sigma^*_c}}{q^2 - m^2_{\Sigma^*_c} + i m_{\Sigma^*_c} \Gamma_{\Sigma^*_c}}(-g^{\mu \nu} +
\frac{\gamma^\mu \gamma^\nu}{3} + \frac{\gamma^\mu q^\nu - \gamma^\nu
q^\mu}{3q^2/m_{\Sigma^*_c}} + \frac{2 q^\mu q^\nu}{3 q^2}).
\end{eqnarray}

\section{Nonrelativistic formalism}
\label{app:nr}

In this appendix, we will describe a nonrelativistic formalism which can be used
to calculate the one-pion-exchange loop diagrams for the decays of the $\bar
D^*\Sigma_c$ and $\bar D\Sigma_c^*$ into the $\bar D^*\Lambda_c$. The reason is
that for these two decays, all the involved particles except for the pion can be
treated nonrelativistically: the $P_c$ is near the $\bar
D^*\Sigma_c$ and $\bar D\Sigma_c^*$ thresholds, and the center-of-mass momentum
in the $\bar D^*\Lambda_c$ system is only 0.43~GeV for $M_{P_c}=4.38$~GeV.

We will take the two-component notation for fields containing heavy
quarks~\cite{Hu:2005gf}. Then the field for charmed mesons is given by $H_a =
\vec D_a^*\cdot \vec\sigma+D_a$, where $D_a$ and $D^*_a$ annihilate the
pseudoscalar and vector charmed mesons, respectively, $a$ is the light flavor
index, and $\vec\sigma$ are the Pauli matrices acting in the spinor space. The
field for anti-charmed mesons reads $\bar H_a = -\vec{\bar{D}}^*_a\cdot
\vec\sigma + \bar D_a$ if we take the phase convention for charge conjugation as
$(D_a,D_a^*)\overset{C}{\to} (\bar D_a,\bar D_a^*)$.
The axial coupling of the pions to the heavy mesons is contained in the
following leading order chiral Lagrangian~\cite{Hu:2005gf,Fleming:2008yn}
\begin{equation}
  \mathcal L_{HH\pi} = - \frac{g}{2} \left\langle H_a^\dag H_b \vec\sigma\cdot
  \vec u_{ba} \right\rangle + \frac{g}{2} \left\langle \bar H_a^\dag \vec\sigma\cdot
  \vec u_{ab} \bar H_b \right\rangle ��
\end{equation}
where $\vec u_{ab} = -\sqrt{2}\vec \partial \phi_{ab}/F + \mathcal{O}(\phi^3)$
contains the pion fields
\begin{equation}
   \phi = \begin{pmatrix}
    \frac1{\sqrt{2}}\pi^0+\frac1{\sqrt{6}}\eta & \pi^+ & K^+ \\ \pi^- &
    -\frac1{\sqrt{2}}\pi^0 + \frac1{\sqrt{6}}\eta & K^0 \\
    K^- & \bar K^0 & -\frac2{\sqrt{6}}\eta
   \end{pmatrix}
\end{equation}
when we consider three light flavors, and $F$ is the pion decay
constant in the chiral limit and we will take 92.2~MeV.
The value of the coupling constant $g=0.57$ can be extracted from the measured
decay width of $D^{*+}$~\cite{Agashe:2014kda}. We can also write down
the trace formalism for the sextet heavy baryon fields introduced in
Ref.~\cite{Georgi:1990cx} in the two-component notation as
\begin{equation}
 S^i_{ab} = B_{6,ab}^{*i} + \frac1{\sqrt{3}} \sigma^i B_{6,ab} ,
\end{equation}
where $B_{6,ab}^{*}$ and $B_{6,ab}$ annihilate the $J^P=\frac32^+$ and
$\frac12^+$ sextet charmed baryons, which degenerate in the heavy quark limit,
respectively. The leading order chiral Lagrangian for the axial pionic
coupling between the sextet and anti-triplet charmed baryons~\cite{Cho:1992gg}
can be written as
\begin{equation}
  \mathcal{L}_{SB_{\bar 3}\pi} = -\frac{\sqrt{3}}{2} g_2 B_{\bar 3,ab}^\dag \vec
  u_{bc}\cdot \vec S_{ca} + \text{h.c.} .
\end{equation}
The charmed baryon matrices in SU(3) flavor space are given by
\begin{equation}
B_{\bar 3} = \begin{pmatrix}
0 & \Lambda_c^+ & \Xi_c^+ \\ - \Lambda_c^+ & 0 & \Xi_c^0 \\ -\Xi_c^+ & -\Xi_c^0
& 0
\end{pmatrix}, 
B_{6} = \begin{pmatrix}
\Sigma_c^{++} & \frac1{\sqrt{2}}\Sigma_c^+ & \frac1{\sqrt{2}}\Xi_c^{\prime+} \\
\frac1{\sqrt{2}}\Sigma_c^+ & \Sigma_c^0 & \frac1{\sqrt{2}}\Xi_c^{\prime0} \\
\frac1{\sqrt{2}}\Xi_c^{\prime+} & \frac1{\sqrt{2}}\Xi_c^{\prime0} & \Omega_c^0
\end{pmatrix} .
\end{equation}
The value of $g_2$ extracted from the decays $\Sigma_c^{*++}\to\Lambda_c^+\pi^+$
and $\Sigma_c^{++}\to\Lambda_c^+\pi^+$ are 0.56 and 0.55, respectively.
At leading order of the nonrelativistic expansion,
summing over the polarizations of the vector meson is given by
\begin{equation}
 \sum_\alpha \epsilon^i(\vec p,\lambda) \epsilon^j(\vec p,\lambda) = \delta^{ij}
 , \label{eq:polvec}
\end{equation}
and summing over the polarizations for
spin-$\frac12$ and spin-$\frac32$ spinors results in
\begin{eqnarray}
\sum_{\alpha} u(\vec p,\alpha) u^{\dag}(\vec p,\alpha) &=& 2 m ,
\nonumber\\
\sum_{\alpha} u^i(\vec p,\alpha) u^{j\dag}(\vec p,\alpha) &=& \frac{2
m}{3} \left( 2 \delta^{ij} - i \varepsilon^{ijk}\sigma^k \right) .
\label{eq:polspin}
\end{eqnarray}

We
define the $S$-wave couplings of the $J^P=\frac32^-$ $P_c$ state to the $\bar
D^*\Sigma_c$ and $\bar D\Sigma_c^*$ by
\begin{equation}
 \mathcal{L}_{P_c} = -\sqrt{\frac23} \left(g_{P_c}^{} \bar D_a^\dag
 \vec\Sigma_{c,ab}^{*\dag}\cdot \vec P_{c,b}
 + g_{P_c}^{\prime} \bar D_a^{*i\dag}
 \Sigma_{c,ab}^{\dag} P_{c,b}^i \right).
\end{equation}
If we assume that the $P_c$ is a $\bar D\Sigma_c^*$ ($\bar D^*\Sigma_c$) bound
state, then $g_{P_c}^2$ ($g_{P_c}^{\prime\,2}$) is given by Eq.~\eqref{eq:coco}
and $g_{P_c}'=0$ ($g_{P_c}=0$).

Using the above Lagrangians, we can calculate the amplitudes for the
one-pion-exchange diagrams shown in Fig.~\ref{Fig:feyndiag}. The
amplitude of $P_c^+(p)\to \Lambda_c^+(k) \bar D^{*0}(q)$ for $P_c$
being the  $\bar D\Sigma_c^*$ molecule and the  $\bar D^*\Sigma_c$
molecule are

\begin{eqnarray}
\mathcal{A}_{\bar D\Sigma_c^*}^\text{OPE} &=& -3N g_{P_c}^{}
m_{\Sigma_c^*}^{} u^\dag(\vec k,\omega)(2\delta^{ij}- i
\varepsilon^{ijk}\sigma^k) u^j(\vec p,\alpha)\epsilon^n(\vec q,\lambda)
I^{in}(m_D,m_{\Sigma_c^*},m_\pi,\vec q\,) , \nonumber\\
\mathcal{A}_{\bar D^*\Sigma_c}^\text{OPE} &=& -i3\sqrt{3}N g_{P_c}^{\prime}
m_{\Sigma_c}^{} \varepsilon^{ijk}
\epsilon^j(\vec q,\lambda) u^\dag(\vec k,\omega) \sigma^n
u^k(\vec p,\alpha)
I^{in}(m_{D^*},m_{\Sigma_c},m_\pi,\vec q\,) ,
\end{eqnarray}
respectively. Here the factor 3 takes into account the contributions from the
isospin multiplets of the intermediate states in the triangle diagrams, $N=2
g g_2^{} g_{P_c}^{}/(3\sqrt{2}F^2)$,
$\omega,\alpha$ and $\lambda$ denote the polarization of the relevant
particles, and the tensor loop integral is defined in the $P_c$
rest frame as
\begin{equation}
  I^{ij}(m_1,m_2,m_3,\vec q\,) \equiv \frac{i}{4 m_1 m_2}
  \int\frac{d^4l}{(2\pi)^4} \frac{l^i l^j}{\left(q^0-l^0-\omega_1+i\epsilon\right)
  \left(k^0+l^0-\omega_2+i\epsilon\right) \left(l^2-m_3^2+i\epsilon\right)} ,
  \label{eq:loop}
\end{equation}
where $\omega_1 = \sqrt{m_1^2+(\vec l - \vec q\,)^2}$, $\omega_2 =
\sqrt{m_2^2+(\vec l - \vec q\,)^2}$, and the propagators of the charmed meson
and baryon have been treated nonrelativistically. The $J^P=\frac32^-$ $P_c$ can
decay into the $\Lambda_c\bar D^*$ in both $S$ wave and $D$ wave. It is
reflected by the fact that the tensor loop integral defined in
Eq.~\eqref{eq:loop} can be decomposed into an $S$-wave part and a $D$-wave part
which will be denoted by $I_S$ and $I_D$, respectively. The decomposition can be
easily done by defining the $S$-wave and $D$-wave projectors
\begin{eqnarray}
P_S^{ij} \equiv \frac13\delta^{ij}, \quad
P_D^{ij} \equiv \frac{q_iq_j}{\vec q^{\,2}}-\frac13\delta^{ij} ,
\end{eqnarray}
which satisfy
$P_S^{ij}P_S^{ij} = 1/3$, $P_D^{ij}P_D^{ij} = 2/3$, and $P_S^{ij}P_D^{ij} = 0$.
We get
\begin{equation}
I_S(m_1,m_2,m_3,\vec q^{\,2}) = I^{ii}(m_1,m_2,m_3,\vec q\,),
\end{equation}
and
\begin{equation}
I_D(m_1,m_2,m_3,\vec q^{\,2}) = \frac32 I^{ij}(m_1,m_2,m_3,\vec q\,) P_D^{ij}.
\end{equation}
It turns out that the $D$-wave part $I_D$ is UV convergent as
discussed in Ref.~\cite{Albaladejo:2015dsa}, while the $S$-wave part $I_S$ is UV
divergent. The divergence might be regularized by introducing a Gaussian form
factor $\exp\left( - (\vec q - \vec l\,)^2/\Lambda^2\right)$, which is the
nonrelativistic analogue of the form factor in Eq.~\eqref{corr_fun}.
For the purpose of qualitatively comparing the relative size of the partial
widths calculated from the one-pion-exchange diagrams for the two
hadronic molecular assignments, we do not need to specify a value for $\Lambda$
as can be seen in the following. Using Eqs.~\eqref{eq:polvec} and
\eqref{eq:polspin}, we have
\begin{eqnarray}
\sum_{\omega,\alpha,\lambda} \left| \mathcal{A}_{\bar
D\Sigma_c^*}^\text{OPE}\right|^2 \al=\al 144 N^2 g_{P_c}^2 m_{\Lambda_c}^{}
m_{P_c}^{} m_{\Sigma_c^*}^2
\Big[ \left|2I_D(m_D,m_{\Sigma_c^*},m_\pi,\vec q^{\,2})\right|^2
+\left|I_S(m_D,m_{\Sigma_c^*},m_\pi,\vec q^{\,2})\right|^2
\Big],\nonumber \\ \label{eq:asq1}\\
\sum_{\omega,\alpha,\lambda} \left| \mathcal{A}_{\bar
D^*\Sigma_c}^\text{OPE}\right|^2 \al=\al 48 N^2 g_{P_c}^{\prime\,2}
m_{\Lambda_c}^{} m_{P_c}^{} m_{\Sigma_c}^2
\Big[5\left|I_D(m_{D^*},m_{\Sigma_c},m_\pi,\vec q^{\,2})\right|^2
+\left|I_S(m_{D^*},m_{\Sigma_c},m_\pi,\vec q^{\,2})\right|^2\Big]
.
\nonumber \\\label{eq:asq2}
\end{eqnarray}
One sees that the partial widths for both the $\bar D\Sigma_c^*$ and $\bar
D^*\Sigma_c$ molecules decays into $\Lambda_c\bar D^*$ contains both $S$-wave
and $D$-wave components, but the $S$-wave component for the decay of the $\bar
D^* \Sigma_c$ molecule is parametrically three times larger than that for the
decay of the $D^*\Sigma_c$. Numerically, the difference is a factor of
$\order{10}$, for $\Lambda$ taking a value in the range of $[0.5,2]$~GeV, after
we have included the kinematic effects that the $P_c(4380)$ mass is closer to
the $\bar D\Sigma_c^*$ threshold than to the $\bar D^*\Sigma_c$ one and that the $\bar D\pi\Lambda_c$ can be on shell simultaneously, if we take
the mass of the $P_c$ to be 4.38~GeV.

Therefore, it is a firm conclusion that if  the $P_c(4380)$ is a $\bar D
\Sigma_c^*$ molecule, its partial width of the decay into $\Lambda_c\bar D^*$ is
much larger than that for the case if it is a $\bar D^*\Sigma_c$ molecule.

\end{appendix}

\bibliographystyle{plain}

\begin{thebibliography}{999}
\bibitem{Aaij:2015tga}
  R.~Aaij {\it et al.} [LHCb Collaboration],
  Phys.\ Rev.\ Lett.\  {\bf 115} (2015) 072001
  doi:10.1103/PhysRevLett.115.072001
  [arXiv:1507.03414 [hep-ex]].

\bibitem{Wu:2010jy}
  J.~J.~Wu, R.~Molina, E.~Oset and B.~S.~Zou,
  Phys.\ Rev.\ Lett.\  {\bf 105} (2010) 232001
  doi:10.1103/PhysRevLett.105.232001
  [arXiv:1007.0573 [nucl-th]]; Phys.\ Rev.\ C {\bf 84} (2011) 015202
  doi:10.1103/PhysRevC.84.015202
  [arXiv:1011.2399 [nucl-th]].


\bibitem{Yuan:2012wz}
  S.~G.~Yuan, K.~W.~Wei, J.~He, H.~S.~Xu and B.~S.~Zou,
  Eur.\ Phys.\ J.\ A {\bf 48} (2012) 61
  doi:10.1140/epja/i2012-12061-2
  [arXiv:1201.0807 [nucl-th]].


\bibitem{Yang:2011wz}
  Z.~C.~Yang, Z.~F.~Sun, J.~He, X.~Liu and S.~L.~Zhu,
  Chin.\ Phys.\ C {\bf 36} (2012) 6
  doi:10.1088/1674-1137/36/1/002
  [arXiv:1105.2901 [hep-ph]].


\bibitem{Xiao:2013yca}
  C.~W.~Xiao, J.~Nieves and E.~Oset,
  Phys.\ Rev.\ D {\bf 88} (2013) 056012
  doi:10.1103/PhysRevD.88.056012
  [arXiv:1304.5368 [hep-ph]].


\bibitem{Uchino:2015uha}
  T.~Uchino, W.~H.~Liang and E.~Oset,
  Eur.\ Phys.\ J.\ A {\bf 52} (2016) no.3,  43
  doi:10.1140/epja/i2016-16043-0
  [arXiv:1504.05726 [hep-ph]].


\bibitem{Chen:2015loa}
  R.~Chen, X.~Liu, X.~Q.~Li and S.~L.~Zhu,
  Phys.\ Rev.\ Lett.\  {\bf 115} (2015) no.13,  132002
  doi:10.1103/PhysRevLett.115.132002
  [arXiv:1507.03704 [hep-ph]].


\bibitem{Chen:2015moa}
  H.~X.~Chen, W.~Chen, X.~Liu, T.~G.~Steele and S.~L.~Zhu,
  Phys.\ Rev.\ Lett.\  {\bf 115} (2015) no.17,  172001
  doi:10.1103/PhysRevLett.115.172001
  [arXiv:1507.03717 [hep-ph]].


\bibitem{Roca:2015dva}
  L.~Roca, J.~Nieves and E.~Oset,
  Phys.\ Rev.\ D {\bf 92} (2015) no.9,  094003
  doi:10.1103/PhysRevD.92.094003
  [arXiv:1507.04249 [hep-ph]].


\bibitem{Mironov:2015ica}
  A.~Mironov and A.~Morozov,
  JETP Lett.\  {\bf 102} (2015) no.5,  271
  doi:10.7868/S0370274X15170038, 10.1134/S0021364015170099
  [arXiv:1507.04694 [hep-ph]].


\bibitem{He:2015cea}
  J.~He,
  Phys.\ Lett.\ B {\bf 753} (2016) 547
  doi:10.1016/j.physletb.2015.12.071
  [arXiv:1507.05200 [hep-ph]].


\bibitem{Huang:2015uda}
  H.~Huang, C.~Deng, J.~Ping and F.~Wang,
  arXiv:1510.04648 [hep-ph].


\bibitem{Burns:2015dwa}
  T.~J.~Burns,
  Eur.\ Phys.\ J.\ A {\bf 51} (2015) no.11,  152
  doi:10.1140/epja/i2015-15152-6
  [arXiv:1509.02460 [hep-ph]].


\bibitem{Guo:2015umn}
  F.~K.~Guo, U.~G.~Mei\ss ner, W.~Wang and Z.~Yang,
  Phys.\ Rev.\ D {\bf 92} (2015) no.7,  071502
  doi:10.1103/PhysRevD.92.071502
  [arXiv:1507.04950 [hep-ph]].


\bibitem{Meissner:2015mza}
  U.~G.~Mei\ss ner and J.~A.~Oller,
  Phys.\ Lett.\ B {\bf 751} (2015) 59
  doi:10.1016/j.physletb.2015.10.015
  [arXiv:1507.07478 [hep-ph]].


\bibitem{Liu:2015fea}
  X.~H.~Liu, Q.~Wang and Q.~Zhao,
  arXiv:1507.05359 [hep-ph].


\bibitem{Lebed:2015tna}
  R.~F.~Lebed,
  Phys.\ Lett.\ B {\bf 749} (2015) 454
  doi:10.1016/j.physletb.2015.08.032
  [arXiv:1507.05867 [hep-ph]].


\bibitem{Wang:2015epa}
  Z.~G.~Wang,
  Eur.\ Phys.\ J.\ C {\bf 76} (2016) no.2,  70
  doi:10.1140/epjc/s10052-016-3920-4
  [arXiv:1508.01468 [hep-ph]].


\bibitem{Scoccola:2015nia}
  N.~N.~Scoccola, D.~O.~Riska and M.~Rho,
  Phys.\ Rev.\ D {\bf 92} (2015) no.5,  051501
  doi:10.1103/PhysRevD.92.051501
  [arXiv:1508.01172 [hep-ph]].


\bibitem{Zhu:2015bba}
  R.~Zhu and C.~F.~Qiao,
  arXiv:1510.08693 [hep-ph].

\bibitem{Shimizu:2016rrd}
  Y.~Shimizu, D.~Suenaga and M.~Harada,
  arXiv:1603.02376 [hep-ph].

\bibitem{Roca:2016tdh}
  L.~Roca and E.~Oset,
  arXiv:1602.06791 [hep-ph].


\bibitem{Molina:2008jw}
  R.~Molina, D.~Nicmorus and E.~Oset,
  Phys.\ Rev.\ D {\bf 78} (2008) 114018
  doi:10.1103/PhysRevD.78.114018
  [arXiv:0809.2233 [hep-ph]].


\bibitem{MuellerGroeling:1990cw}
  A.~Mueller- Groeling, K.~Holinde and J.~Speth,
  Nucl.\ Phys.\ A {\bf 513} (1990) 557.
  doi:10.1016/0375-9474(90)90398-6


\bibitem{Chen:2013bha}
  D.~Y.~Chen, X.~Liu and T.~Matsuki,
  Phys.\ Rev.\ D {\bf 88} (2013) no.1,  014034
  doi:10.1103/PhysRevD.88.014034
  [arXiv:1306.2080 [hep-ph]].

\bibitem{Oh:2000qr}
  Y.~S.~Oh, T.~Song and S.~H.~Lee,
  Phys.\ Rev.\ C {\bf 63} (2001) 034901
  doi:10.1103/PhysRevC.63.034901
  [nucl-th/0010064].

\bibitem{Oh:2007jd}
  Y.~Oh, C.~M.~Ko and K.~Nakayama,
  Phys.\ Rev.\ C {\bf 77} (2008) 045204
  doi:10.1103/PhysRevC.77.045204
  [arXiv:0712.4285 [nucl-th]].

\bibitem{Can:2012tx}
  K.~U.~Can, G.~Erkol, M.~Oka, A.~Ozpineci and T.~T.~Takahashi,
  Phys.\ Lett.\ B {\bf 719} (2013) 103
  doi:10.1016/j.physletb.2012.12.050
  [arXiv:1210.0869 [hep-lat]].

\bibitem{Ronchen:2012eg}
  D.~Ronchen {\it et al.},
  Eur.\ Phys.\ J.\ A {\bf 49} (2013) 44
  doi:10.1140/epja/i2013-13044-5
  [arXiv:1211.6998 [nucl-th]].

\bibitem{Aliev:2011kn}
  T.~M.~Aliev, K.~Azizi, M.~Savci and V.~S.~Zamiralov,
  Phys.\ Rev.\ D {\bf 83} (2011) 096007
  doi:10.1103/PhysRevD.83.096007
  [arXiv:1102.0416 [hep-ph]].

\bibitem{Aliev:2010ev}
  T.~M.~Aliev, K.~Azizi and M.~Savci,
  Eur.\ Phys.\ J.\ C {\bf 71} (2011) 1675
  doi:10.1140/epjc/s10052-011-1675-5
  [arXiv:1012.5935 [hep-ph]].

\bibitem{Aliev:2010yx}
  T.~M.~Aliev, K.~Azizi and M.~Savci,
  Phys.\ Lett.\ B {\bf 696} (2011) 220
  doi:10.1016/j.physletb.2010.12.027
  [arXiv:1009.3658 [hep-ph]].

\bibitem{Aliev:2010su}
  T.~M.~Aliev, K.~Azizi and M.~Savci,
  Nucl.\ Phys.\ A {\bf 847} (2010) 101
  doi:10.1016/j.nuclphysa.2010.06.013
  [arXiv:1003.5467 [hep-ph]].

\bibitem{Garzon:2015zva}
  E.~J.~Garzon and J.~J.~Xie,
  Phys.\ Rev.\ C {\bf 92} (2015) no.3,  035201
  doi:10.1103/PhysRevC.92.035201
  [arXiv:1506.06834 [hep-ph]].

\bibitem{Guo:2010ak}
  F.~K.~Guo, C.~Hanhart, G.~Li, U.~G.~Mei\ss ner and Q.~Zhao,
  Phys.\ Rev.\ D {\bf 83} (2011) 034013
  doi:10.1103/PhysRevD.83.034013
  [arXiv:1008.3632 [hep-ph]].

\bibitem{Gao:2010ve}
  P.~Gao, B.~S.~Zou and A.~Sibirtsev,
  Nucl.\ Phys.\ A {\bf 867} (2011) 41
  doi:10.1016/j.nuclphysa.2011.07.008
  [arXiv:1011.2387 [nucl-th]].

\bibitem{Kim:2015ita}
  S.~H.~Kim, A.~Hosaka, H.~C.~Kim and H.~Noumi,
  Phys.\ Rev.\ D {\bf 92} (2015) no.9,  094021
  doi:10.1103/PhysRevD.92.094021
  [arXiv:1509.03567 [hep-ph]].


\bibitem{Zou:2002yy}
  B.~S.~Zou and F.~Hussain,
  Phys.\ Rev.\ C {\bf 67} (2003) 015204
  doi:10.1103/PhysRevC.67.015204
  [hep-ph/0210164].


\bibitem{Faessler:2008zza}
  A.~Faessler, T.~Gutsche and V.~E.~Lyubovitskij,
  Prog.\ Part.\ Nucl.\ Phys.\  {\bf 61} (2008) 127.
  doi:10.1016/j.ppnp.2007.12.005


\bibitem{Chen:2015igx}
  D.~Y.~Chen and Y.~B.~Dong,
  Phys.\ Rev.\ D {\bf 93} (2016) no.1,  014003
  doi:10.1103/PhysRevD.93.014003
  [arXiv:1510.00829 [hep-ph]].


\bibitem{Lu:2016nnt}
  Q.~F.~L\"u and Y.~B.~Dong,
  arXiv:1603.00559 [hep-ph].


\bibitem{Huang:2013mua}
  Y.~Huang, J.~He, H.~F.~Zhang and X.~R.~Chen,
  J.\ Phys.\ G {\bf 41} (2014) no.11,  115004
  doi:10.1088/0954-3899/41/11/115004
  [arXiv:1305.4434 [nucl-th]].


\bibitem{Lu:2015fva}
  Q.~F.~L\"u, X.~Y.~Wang, J.~J.~Xie, X.~R.~Chen and Y.~B.~Dong,
  Phys.\ Rev.\ D {\bf 93} (2016) no.3,  034009
  doi:10.1103/PhysRevD.93.034009
  [arXiv:1510.06271 [hep-ph]].


\bibitem{Weinberg:1965zz}
  S.~Weinberg,
  Phys.\ Rev.\  {\bf 137} (1965) B672.
  doi:10.1103/PhysRev.137.B672


\bibitem{Baru:2003qq}
  V.~Baru, J.~Haidenbauer, C.~Hanhart, Y.~Kalashnikova and A.~E.~Kudryavtsev,
  Phys.\ Lett.\ B {\bf 586} (2004) 53
  doi:10.1016/j.physletb.2004.01.088
  [hep-ph/0308129].


\bibitem{Guo:2008zg}
  F.~K.~Guo, C.~Hanhart and U.~G.~Meissner,
  Phys.\ Lett.\ B {\bf 665} (2008) 26
  doi:10.1016/j.physletb.2008.05.057
  [arXiv:0803.1392 [hep-ph]].


\bibitem{Hu:2005gf}
  J.~Hu and T.~Mehen,
  Phys.\ Rev.\ D {\bf 73} (2006) 054003
  doi:10.1103/PhysRevD.73.054003
  [hep-ph/0511321].


\bibitem{Fleming:2008yn}
  S.~Fleming and T.~Mehen,
  Phys.\ Rev.\ D {\bf 78} (2008) 094019
  doi:10.1103/PhysRevD.78.094019
  [arXiv:0807.2674 [hep-ph]].


\bibitem{Agashe:2014kda}
  K.~A.~Olive {\it et al.} [Particle Data Group Collaboration],
  Chin.\ Phys.\ C {\bf 38} (2014) 090001.
  doi:10.1088/1674-1137/38/9/090001


\bibitem{Georgi:1990cx}
  H.~Georgi,
  Nucl.\ Phys.\ B {\bf 348} (1991) 293.
  doi:10.1016/0550-3213(91)90519-4


\bibitem{Cho:1992gg}
  P.~L.~Cho,
  Phys.\ Lett.\ B {\bf 285} (1992) 145
  doi:10.1016/0370-2693(92)91314-Y
  [hep-ph/9203225].


\bibitem{Albaladejo:2015dsa}
  M.~Albaladejo, F.-K.~Guo, C.~Hidalgo-Duque, J.~Nieves and M.~P.~Valderrama,
  Eur.\ Phys.\ J.\ C {\bf 75} (2015) no.11,  547
  doi:10.1140/epjc/s10052-015-3753-6
  [arXiv:1504.00861 [hep-ph]].

\end{thebibliography}

\end{document}